\documentstyle[12pt]{article}

\newcommand \beq{\begin{eqnarray}}
\newcommand \eeq{\end{eqnarray}}
\def\bfgrad{\mbox{\boldmath$\grad$}}

\def\grad{\nabla}                               
\def\del{\partial}                              

\def\frac#1#2{{#1 \over #2}}

\def\half{\ifinner {\scriptstyle {1 \over 2}}
   \else {1 \over 2} \fi}

\def\simge{\mathrel{%
   \rlap{\raise 0.511ex \hbox{$>$}}{\lower 0.511ex \hbox{$\sim$}}}}
\def\simle{\mathrel{
   \rlap{\raise 0.511ex \hbox{$<$}}{\lower 0.511ex \hbox{$\sim$}}}}

\def\slashchar#1{\setbox0=\hbox{$#1$}           
   \dimen0=\wd0                                 
   \setbox1=\hbox{/} \dimen1=\wd1               
   \ifdim\dimen0>\dimen1                        
      \rlap{\hbox to \dimen0{\hfil/\hfil}}      
      #1                                        
   \else                                        
      \rlap{\hbox to \dimen1{\hfil$#1$\hfil}}   
      /                                         
   \fi}                                         %

\def\subrightarrow#1{
  \setbox0=\hbox{
    $\displaystyle\mathop{}
    \limits_{#1}$}
  \dimen0=\wd0
  \advance \dimen0 by .5em
  \mathrel{
    \mathop{\hbox to \dimen0{\rightarrowfill}}
       \limits_{#1}}}                           

\def\journal#1#2#3#4{\ {#1}{\bf #2} ({#3})\  {#4}}

\def\NPB{\journal{Nucl.\ Phys.\ {\bf B}}}

\def\PRD{\journal{Phys.\ Rev.\ {\bf D}}}
\def\PRL{\journal{Phys.\ Rev.\ Lett.}}
\def\PhysRept{\journal{Phys.\ Repts.}}

\begin{document}

\begin{titlepage}
\begin{flushright}
{Saclay-T94/041}
\end{flushright}
\vspace*{2cm}
\begin{center}
\baselineskip=13pt
{\Large  GAUGE STRUCTURE AND SEMI-CLASSICAL ASPECTS OF HARD THERMAL
LOOPS.\\}
\vskip0.5cm
Jean-Paul BLAIZOT\footnote{CNRS}  and
Edmond IANCU\\
{\it Service de Physique Th\'eorique\footnote{Laboratoire de la Direction
des
Sciences de la Mati\`ere du Commissariat \`a l'Energie
Atomique}, CE-Saclay \\ 91191 Gif-sur-Yvette, France}\\
\vskip0.5cm
April 1994
\end{center}

\vskip 2cm
\begin{abstract}
Hard thermal loops play a central role in the theory of long
wavelength excitations of a quark-gluon plasma. We show in this paper how
our
recent derivation of their generating functional from the Dyson-Schwinger
equations sheds light on their semi-classical nature and their remarkable gauge
structure. In particular, we show that our kinetic equations
can be written in terms of auxiliary gauge fields with zero curvature.
Remarkably, the latter property determines entirely the dynamics in the
kinematical regime of hard thermal loops. This explains in particular why the
generating functional could have been obtained by Taylor and Wong from gauge
invariance considerations. It also clarifies the role of Chern-Simons theory
in this context.
 \end{abstract}

\vskip 3cm


\end{titlepage}

\section{Introduction}

``Hard thermal loops'' (HTL) are the dominant contributions at high temperature
to $n$-point amplitudes with soft external momenta ($P\sim gT$, with $T$ the
temperature and $g$ the gauge coupling, assumed to be small). Their role in
perturbative calculations, and their remarkable properties, where first
established via a systematic diagrammatic analysis\cite{Pisarski,BP90,FT90}.
One of their most
noticeable property is their gauge symmetry; specifically, hard thermal loops
obey Ward identities similar to those of the tree amplitudes,
suggesting that their generating functional should be a gauge invariant
functional of the soft fields. The
gauge invariance has been used by Taylor and Wong to construct a generating
functional for the HTL's in closed form\cite{TW90}.
Later, this effective action has been rewritten in a manifestly gauge
invariant form by Braaten and Pisarski\cite{PB92} and also by
Frenkel and Taylor\cite{FT92}.  It was also realized that  the  generating
functional of HTL's could be understood as the eikonal of a Chern-Simons gauge
 theory, raising the hope that
the use of the known properties of the latter theory could lead to a better
understanding of gauge theories at finite temperature\cite{EN92}.

In a series of recent papers, we have shown that the hard thermal loops are
intimately related to the long wavelength, collective, excitations of the
hot gauge plasmas, much alike the non trivial dielectric properties of the
electromagnetic plamas are related to the collective oscillations of their
charge constituents. To establish this correspondence, we have developed a
 theory of the collective excitations, which is gauge covariant and
takes consistently
 into account all the leading terms in an expansion of the equations of
motion in power of the coupling strength\cite{QED,prl1,NPB2}.

A central quantity in our approach is the ``induced current'' which represents
the response of the plasma particles to applied background gauge fields.
This is obtained, after solving nonlinear equations which may be given the
form of kinetic equations, as a non local functional of the gauge fields. The
nonlinearity of these equations
 is entirely due to gauge rotations of the current (a vector in
color space) induced by covariant derivatives. The non locality of the
induced current arises from
the need, in calculating the response functions, to integrate the history of
the hard particles  along their straight line
 trajectories. These simple features, whose physical origin is tranparent,
explain most of the properties of the HTL's, properties which remained
somewhat surprising in the diagrammatic approach.

In this paper we wish to pursue our analysis, and make closer contact with
the formal works mentioned above. In particular we shall show that the gauge
structure of the HTL's follows from a simple property of our kinetic
equations, which is ultimately related to a property of the Lorentz equation
describing the
motion of a classical particle in an electromagnetic field, or its
generalisation to the non abelian case. More precisely, we shall show that,
in the same way that the Lorentz equation
 can be rewritten in terms of the gauge dependent {\it
canonical} momentum, our kinetic equations can be also written in terms of
auxiliary gauge potentials with {\it zero field strength}. The
latter property follows from the fact that, within the present
approximation scheme, the plasma particles do not change direction in the
course of their interaction with the gauge field. Quite remarkably,
in this kinematical regime, the
condition of zero field strength turns out to be sufficient to determine
the dynamics. As we shall see, this provides an explanation for the fact
that it has been possible to construct the generating functional for the hard
thermal loops on the basis of gauge invariance only, with no dynamical
input beyond the form of the gluon self energy\cite{TW90,PB92}.

The plan of the paper is as follows. In the next section, we present a
short summary of our approach, and recall the form of our basic kinetic
equations. In section 3 we discuss some of their semi-classical
features, and show that they can be given a simple gauge structure. Then in
section 4 we show explicitely how the gauge invariance condition used by
Taylor and Wong to derive the generating functional of HTL follows from our
kinetic equations. We also clarify the role of Chern-Simons theory whose
relevance in the present context has been much advocated recently.  Finally,
we verify that the expression for the energy density obtained by Nair using
Chern-Simons theory coincides with the one that we have obtained by
integrating the mean field equations. The last  section summarizes our
conclusions.

\setcounter{equation}{0}
\section{Semiclassical theory of collective excitations}

Our semi-classical theory of the  collective excitations is based on a
 perturbative analysis of the QCD Dyson-Schwinger equations.
 It is obtained by consistently preserving the leading terms in an expansion
in powers of the gauge coupling $g$. In doing so, one encounters three types
of approximations, which, in most many-body systems, are
independent approximations, but here are controlled
 by the same small parameter, i.e. $g$. These are the weak coupling
 approximation ($g\ll 1$), the
 long wavelength approximation ($\lambda\sim 1/gT\gg 1/T$), and the small
amplitude approximation (the gauge field strength tensor is limited by
 $F\simle gT^2$, or, equivalently, the gauge potentials satisfy $A\simle T$).

The result of our analysis is a set of coupled equations.
  The first equation, the
generalization of the Maxwell equation in a polarizable medium,
relates  the gauge mean field $A_\mu^a$ to  the induced current $j^a_\mu$
(see (a) below).
The other equations allow the calculation of the induced current
in terms of the average gauge fields. They can be given the form of simple
kinetic equations, analogous to Vlasov's equation of ordinary plasma
physics (see (c) below). In our approach, the hard thermal loops appear
as functional derivatives of the induced current with respect to the
gauge fields\cite{QED,prl1,NPB2}.

(a) {\it Mean field equation}

The equation of motion for $A_\mu^a$ takes the form
\beq\label{ava}
\left [\, D^\nu,\, F_{\nu\mu}(x)\,\right ]^a
\,=\,j_\mu^a(x),
\eeq
where $D_\mu = \del_\mu+igA_\mu(x),$ ($A_\mu\equiv A_\mu^a t^a$),
  $F_{\mu\nu}= [D_\mu, D_\nu]/(ig)$ is the field strength tensor,
and $j_\mu(x)$ is the induced current, which transforms as a color vector in
the adjoint representation. (The generators of the gauge group in  different
 representations
are taken to be Hermitian and traceless. They are denoted by $t^a$ and
$T^a$, respectively, for the fundamental and the adjoint representation,
and are normalized so that Tr$(t^a t^b)=(1/2)\delta^{ab}$ and
 Tr$(T^aT^b)=N\delta^{ab}$.)

(b) {\it Structure of the induced current}

The induced current expresses the response of the plasma  particles
to the long wavelength color fields $A_\mu^a$.
It is convenient to separate the contributions of the
fermions from those of the transverse gluons, by writing
 $j_\mu^{a}=j_{{\rm f}\,\mu}^{\,a}+j_{{\rm b}\,\mu}^{\,a}$. For the
 fermions (with $N_f$ flavors) we have
\beq\label{jf}
j_{{\rm f}\,\mu}^{\,a}(x)&=&gN_f\int\frac{d^3k}{(2\pi)^3}\,v_\mu
\Bigl( \delta
 n_+^a({\bf k},x)-\delta n_-^a({\bf k},x)\Bigr),\eeq
where $v^\mu\equiv (1,\,{\bf v})$ and ${\bf v}\equiv
{\bf k}/k$ is the velocity of the hard particle ($k\equiv |{\bf k}|$,
and $|{\bf v}| =1$). The quantities $\delta n_\pm^a({\bf k},x)$
are the components of a color vector in the adjoint representation. They may
be combined into a density matrix,
  $\delta n_\pm \equiv\delta n_\pm^a\,t^a$,  ultimately
related to the fermion propagator in the presence of the gauge
fields. There is no color singlet component in   $\delta n_\pm$ because we are
considering only excitations which carry gluon quantum numbers.

Similarly, the current carried by the transverse gluons can be written
in terms of a density matrix
$\delta N({\bf k},x)\equiv \delta N^a({\bf k},x)\,T^a$:
\beq\label{jb}
j_{{\rm b}\,\mu}^{\,a}(x)=2g\int\frac{d^3k}{(2\pi)^3}\,v_\mu
\,{\rm Tr}\,T^a\,\delta N({\bf k},x)=2gN\int\frac{d^3k}{(2\pi)^3}\,v_\mu
\,\delta N^a({\bf k},x).\eeq

(c) {\it Kinetic equations for $\delta n_\pm $ and $\delta N$}

The color density matrices  $\delta n_\pm$ and $\delta N$ are determined
by the following kinetic equations\cite{QED,NPB2}
\beq\label{n}
\left[ v\cdot D_x,\,\delta n_\pm({{\bf k}},x)\right]=\mp\, g\,{\bf v}
\cdot{\bf E}(x)\,\frac{dn_k}{dk},\eeq
\beq\label{N}
\left[ v\cdot D_x,\,\delta N({{\bf k}},x)\right]=-\, g\,
{\bf v}\cdot{\bf E}(x)\frac{dN_k}{dk},\eeq
 where $E_a^i\equiv F_a^{i0}$ is the chromoelectric field. Furthermore,
$N_k \equiv 1/(\exp(\beta k)-1)$ and  $n_k\equiv 1/(\exp(\beta k)+1)$ are,
 respectively, the equilibrium boson and fermion occupation factors.
Note that, in deriving these equations, we have used the isotropy (in
 ${\bf k}$ space) of the equilibrium distribution functions. For
more general, anisotropic, equilibrium distributions
 $n({\bf k})$ and  $N({\bf k})$, we should replace
$${\bf v}\cdot{\bf E}(x)\,\frac{dn_k}{dk} \longrightarrow F^{i \nu}(x)v_\nu
\frac{\del n}{\del k^i},$$
in the r.h.s. of eq.~(\ref{n}), and similarly for the second equation.

In the abelian case, eq.(2.4) reduces to the well known linearized Vlasov
equation describing the evolution of single particle distribution functions.
It is perhaps useful here to emphasize that, although the underlying dynamics
is that of the Lorentz equation describing the motion of a classical charged
particle in an electromagnetic field (see next section), however $\delta n({\bf
k},x)$ is a Wigner function whose only {\it slow} variations in $x$ have a
classical character (in particular, the plasma particles have typical
wavelength comparable to their mean separation and remain therefore
quantum particles). In the non abelian case, the
 distribution functions are matrices
in color space: the color degree of freedom remains essentially quantum. (In
this respect we differ from Heinz who has given a derivation of equations
similar to ours starting from classical kinetic theory, and treating color
also as a classical degree of freedom\cite{Heinz}.)

For the subsequent analysis of these equations, it is convenient to express
 $\delta n_\pm$ and $\delta N$ in terms of  new functions,
 $W^\mu(x;v)\equiv W_a^\mu(x;v) t^a$, defined as the solutions to
\beq\label{eqw}
\left[ v\cdot D_x,\, W^\mu(x;v)\right]\,=\,F^{\mu\nu}(x)\,v_\nu.\eeq
The quantities $W_a^\mu(x;v)$
are the components of a color vector in the adjoint representation, and satisfy
\beq\label{trans} v_\mu\,W^\mu_a(x;v)\,=\,0,\eeq
so that  $W_a^0 =v^iW_a^i$. From the equations above, it is easily seen that
\beq\label{dnN}
\delta n_\pm^a({\bf k}, x)=\mp\, g W^0_a(x;v)\,\frac{dn}{dk},\qquad
\delta N^a({\bf k}, x)=- gW^0_a(x;v)\,\frac{dN}{dk}.\eeq
Note that for an anisotropic initial distribution, the fluctuations
 $\delta n^a_\pm $ and $\delta N^a$ depend on {\it  all} the three independent
components of $W_a^\mu(x;v)$, that is, e.g.,
$$\delta n_\pm^a({\bf k}, x)=\mp\, g W^i_a(x;v)\,\frac{\del n}{\del k^i},$$
and similarly for  $\delta N^a$.

(d) {\it Solving the kinetic equations}

The equation (\ref{eqw}) for $W^\mu_a$ is easily solved once boundary
conditions are given. We assume here retarded conditions, such that
the fields $A_\mu$ vanish as  $x_0\to -\infty$.
The corresponding solution of eq.~(\ref{eqw}) reads
\beq\label{W}
W^\mu(x;v)\,=\, \int_0^\infty d\tau\, U(x,x-v\tau)\, F^{\mu\nu}
(x-v\tau)\,v_\nu\,U(x-v\tau,x),\eeq
where  $U(x,y)$  is the parallel transporter along the straight line
$\gamma$   joining $x$ and $y$,
\beq\label{pt} U(x,y)=P\exp\{ -ig\int_\gamma dz^\mu A_\mu(z)\},\eeq
with $P$ denoting the path-ordering operator. Thus,
\beq\label{U}
U(x,x-v\tau)=P\exp\left\{ -ig\int_0^\tau ds \, v\cdot A(x-v(\tau - s))
\right\}.\eeq
In order to verify that (\ref{W}) is the correct solution to eq.~(\ref{eqw})
 we may use the following formula for the line-derivative of the parallel
transporter
\beq\label{derU}
(v\cdot \del_x)\,U(x,y)\Big |_{y=x-v\tau}= -ig\,v\cdot A(x)\,U(x,x-v\tau).
\eeq
We shall use later the quantity
\beq\label{defV}
V(x;v)\equiv \lim_{\tau\to \infty}\,U(x,x-v\tau),\eeq
which is well defined since  $A_\mu(x-v\tau)\to 0$ as
$\tau\to \infty$. We have
\beq\label{eqV}
(v\cdot \del)\,V(x;v)= -ig\,v\cdot A(x)\,V(x;v),\eeq
which leads to the following representation for $v\cdot A(x)$:
\beq\label{vA}
v\cdot A(x)\,=\,\frac{i}{g}\,\Bigl( (v\cdot \del)V\Bigr)\,V^{-1}.\eeq
The  matrix $V(x;v)$ generates the gauge transformation from
the gauge $v\cdot A=0$ to the actual gauge.

(e) {\it Induced current in closed form}

Once the solution of the kinetic equation is known, one can easily calculate
the  induced current in closed form. By inserting eqs.~(\ref{dnN})
in the expressions (\ref{jf}) and (\ref{jb}), we obtain
\beq\label{j1}
j^\mu_a(x)\,=\,3\,\omega^2_p\int\frac{d\Omega}{4\pi}
\,v^\mu\,W_a^0(x;v).\eeq
 after the integration over $k=|{\bf k}|$ has been performed.
Here $\omega_p$ is the {\it plasma frequency},
$\omega^2_p\equiv (2N+N_{\rm f})g^2 T^2/18$. The  integral $\int
d\Omega$ runs over all the directions of the unit vector ${\bf v}$.
According to  eqs.~(\ref{W}) and (\ref{j1}), the retarded current reads
\beq\label{jind}
j^\mu(x)\,=\,3\,\omega^2_p\int\frac{d\Omega}{4\pi}
\,v^\mu \int_0^\infty d\tau\, U(x,x-v\tau)\, {\bf v}\cdot{\bf E}
(x-v\tau)\,U(x-v\tau,x).\eeq
This expression can be viewed as a generating functional of the hard thermal
loops with retarded boundary conditions. It was first presented
in Ref. \cite{prl1}, and has been rederived since using other
 methods\cite{JN93} (see also below).
 The current (\ref{j1}) is covariantly conserved,
\beq\label{cons}
\Bigl[ D_\mu, j^\mu(x)\Bigr]&=&0,\eeq
as necessary for the consistency of the mean field equation (\ref{ava}).
The  ``abelian-like'' Ward identities
relating the HTL's, which were first identified in the diagrammatic approach
 of Refs. \cite{BP90,FT90}, are a direct consequence of eq.~(\ref{cons}).

\setcounter{equation}{0}
\section{Gauge structure}

We turn now to a physical interpretation of some of the quantities
that we have introduced, and to an analysis of the simple gauge structure
which emerges from our equations.

First, it is useful to  recall  various forms of the
equations of motion for a classical particle of mass $m$ and charge $e$,
 moving in an electromagnetic
 background field. From the Lagrange function\cite{LandauTDC}
\beq\label{L}
L\,=\,-\,m\sqrt{1-{\bf v}^2}\,-\,e\,v\cdot A,\eeq
one easily obtains ($v^\mu\equiv (1, {\bf v})$)
\beq\label{LP}
\frac{d{\bf p}}{dt}\,=\,-e\,v^\mu\,{\bfgrad} A_\mu.\eeq
The {\it canonical} momentum
\beq\label{can}
{\bf p}=\frac{\del L}{\del {\bf v}}=\frac{m{\bf v}}{\sqrt{1-{\bf v}^2}}\,+\,
e{\bf A}\equiv {\bf k}+e{\bf A}\eeq
is the sum of the {\it kinetic} momentum ${\bf k}$, related to the
velocity of the particle, and $e$ times the gauge field ${\bf A}$. The
energy of the particle may be given a similar form,
\beq\label{canE}
p^0={\bf p}\cdot {\bf v} - L =\frac{m}{\sqrt{1-{\bf v}^2}}\,+\,
eA^0\equiv { k}^0+e{A}^0.\eeq
It obeys the equation
\beq\label{Len}
\frac{dp^0}{dt}\,=\,e\,v^\mu\del_0A_\mu,\eeq
so that, for the 4-momentum $p^\mu$, we have:
\beq\label{LE}
\frac{dp^\mu}{dt}\,=\,e\,v^\nu\del^\mu A_\nu.\eeq
This equation is not manifestly gauge covariant.
However, the  equation for the kinetic 4-momentum $k^\mu=p^\mu-eA^\mu$,
\beq\label{lorentz}
\frac{d k^\mu}{dt}\,=\,e\,F^{\mu\nu}(x)\,v_\nu,\eeq
is explicitly independent of the choice of the  gauge. We shall see that,
 to the two ways of writing the Lorentz equation, namely eqs.~(\ref{LE}) and
 (\ref{lorentz}), correspond two ways of writing our basic kinetic equation
(\ref{eqw}) for $W^\mu$.

Consider first  a  QED plasma, where eq.~(\ref{eqw}) reduces to
\beq\label{abw}
( v\cdot \del_x)\, W^\mu(x;v)\,=\,F^{\mu\nu}(x)\,v_\nu.\eeq
This is very much similar to eq.~(\ref{lorentz}).
To make the analogy closer, we
remark that the line derivative in the l.h.s. of eq.~(\ref{abw}), i.e.
 $v\cdot\del_x=\del_t+{\bf v}\cdot {\bfgrad}$,
may be interpreted as the total time derivative along the
 trajectory ${\bf x}(t)={\bf x}_0+{\bf v}t$ of a fictitious particle.
That is, eq.~(\ref{abw}) may be rewritten as
\beq\label{wlor}
\frac{d}{dt}\, W^\mu(t,{\bf x}(t);v)\,=\,F^{\mu\nu}(t,{\bf x}(t))\,v_\nu,\eeq
showing that $e\,dW^\mu(x;v)$ is
the kinetic 4-momentum acquired during the time $dt$ by a charged particle due
to its interaction with the electromagnetic field.
Note, however, that in calculating $W^\mu$ according to eq.~(\ref{W})
we have assumed the velocity
${\bf v}$ to be constant, while in the Lorentz equation
 (\ref{lorentz}), $k^\mu$ and ${\bf v}$ are related quantities
 ($k^\mu=k^0 v^\mu$).
This is in line with our assumption that the typical momenta of
the thermal particles are ``hard'', $k\sim T$, while those of the
background fields are ``soft'', $P\sim gT\ll k$. Thus the hard particles
follow straight line trajectories at the speed of the light, and these
trajectories are not altered, to leading order, by the interaction with the
background gauge field. The condition (\ref{trans}) reflects then simply
the fact that the energy transferred by the
field, $eW^0$, coincides with the mechanical work done by the Lorentz force,
$e{\bf v}\cdot\Delta{\bf k}=e v^i W^i$.
With this interpretation, the formulae (\ref{dnN}) for
 $\delta  n_\pm$ become transparent:
\beq\delta n_\pm({\bf k}, {\bf x}, t)
&=&\mp e W^0\,\frac{dn}{dk}
=\mp e W^i\,\frac{\del n}{\del k^i}=-\Delta k_\pm^i\,\frac{\del n}{\del k^i}
\nonumber\\ &\approx & n \bigl({\bf k}-\Delta{\bf k}_\pm({\bf x},t)\bigr)
\,-\,n({\bf k}),\eeq
that is, the particles found at time $t$ at the point ${\bf x}$ with
momentum ${\bf k}$, are those which, in the initial distribution, have
momentum ${\bf k}-\Delta{\bf k}({\bf x},t)$, where $\Delta{\bf k}({\bf x},t)$
is the momentum acquired by the particles whose trajectories go through
${\bf x}$ at time $t$.

In the non-abelian case,  the fluctuations
$\delta  n_\pm$ and $\delta N$ are matrices in color space, that is,
 $W^\mu=W^\mu_a t^a$. The color
vector of components $W^\mu_a$ precesses
in the background gauge field. This precession
is induced by the covariant derivative in eq.~(\ref{eqw}). Viewing this
precession as an additional source of time-dependence for the  color
vector $W^\mu_a$, one can write
\beq\label{wnab}
\frac{d}{dt}\, W_a^\mu(t,{\bf x}(t);v)\,=\,\Bigl[\bigl(\del_t+{\bf v}\cdot
{\bfgrad}\bigr)\delta_{ac}\,-\,gf_{abc}\,(v\cdot A_b)\Bigr]\,W^\mu_c,\eeq
so that eq.~(\ref{eqw}) may be given a form similar to eq.~(\ref{wlor}).

We shall see now that it is possible to rewrite our kinetic equation
in a form similar to eq.~(\ref{LE}) for the canonical momentum.
This follows by noticing that
\beq
F^{\mu\nu}\,v_\nu\,=\,\del^\mu(v\cdot A)\,-\,\Bigl[v\cdot D, A^\mu\Bigr].\eeq
Thus, if one sets
\beq\label{defa}
a^\mu(x;v)\equiv A^\mu(x)\,+\,W^\mu(x;v),\eeq
one gets\beq\label{eqa}
\bigl[v\cdot D,\, a^\mu\bigr]\,=\,\del^\mu(v\cdot A),\eeq
showing that $a^\mu$ is a functional of $v\cdot A$ only. Explicitly,
\beq\label{a1}
a^\mu(x;v)\,=\, \int_0^\infty d\tau\, U(x,x-v\tau)\, \del^\mu\Bigl(v\cdot
A(x-v\tau)\Bigr)\,U(x-v\tau,x),\eeq
where the retarded boundary conditions have been taken into account.
In the abelian case, eq.~(\ref{eqa}) can be rewritten in a form
 analogous to eq.~(\ref{LE}) for $p^\mu$:
\beq\label{La}
\frac{d}{dt}\, a^\mu(t,{\bf x(t)};v)=(\del_t+{\bf v}\cdot{\bfgrad})\,a^\mu
=\del^\mu(v\cdot A),\eeq
showing that $ga^\mu$ may be indeed  understood as the
change in the {\it canonical} momentum $p^\mu\equiv k^\mu+gA^\mu$
of a fictitious particle following the trajectory ${\bf x}(t)={\bf x}_0
+{\bf v}t$.

The two writings of our kinetic equations, namely eq.~(\ref{eqw}) and
eq.~(\ref{eqa}), have their respective advantages. Eq.~(\ref{eqw}) is
manifestly gauge covariant, while eq.~(\ref{eqa}) is not. On the other
hand, the new fields $a^\mu$ transform as gauge potentials, which has
interesting consequences, as we shall see now.
Under a gauge transformation induced by the operator
 $h(x)=\exp\bigl(-ig\theta^a(x) t^a\bigr)$,
$W^\mu=W^\mu_at^a$ transforms as
$W^\mu(x)\to  h(x)\,W^\mu(x)\,h^{-1}(x)$, while
$A_\mu=  A_\mu^at^a\to   h\,A_\mu\,h^{-1}
\,-\,(i/g)\,h\del_\mu h^{-1}$.
Thus $a^\mu(x;v)$ transforms indeed as a gauge potential. Let
 ${\cal D}_\mu\equiv
\del_\mu+ig\,a_\mu$ be the  covariant derivative  constructed from $a_\mu$, and
$f_{\mu\nu}(x;v)\equiv [{\cal D}_\mu,{\cal D}_\nu]/(ig)$ the corresponding
 field strength tensor. Then it may be shown that
\beq\label{0f}  f_{\mu\nu}=
\del_\mu a_\nu\,-\,\del_\nu a_\mu\,+\,ig\,\bigl[a_\mu, a_\nu\bigr]\,=\,0.\eeq
To verify (\ref{0f}), one could directly derive the equation satisfied by
 $f_{\mu\nu}$,
starting with eq.~(\ref{eqa}). This is the way followed in Ref. \cite{EMT},
where eq.~(\ref{0f}) was first obtained (see eq.~(3.24) in  Ref. \cite{EMT}).
 An alternative derivation is to solve eq.~(\ref{eqa}) for $a^\mu$, with
 $v\cdot A$ given by eq.~(\ref{vA}). It is then easily seen that
that \beq\label{a2}
a^\mu(x;v)\,=\,\frac{i}{g}\,\Bigl(\del^\mu V\Bigr)\,V^{-1},\eeq
from which  eq.~(\ref{0f}) follows.

In summary,  our basic dynamical equation
 can be written either
in terms of $W^\mu$ (see eq.~(\ref{eqw})), or in terms of $a^\mu$
(as in eq.~(\ref{eqa})). In the abelian case, this freedom corresponds
 to the choice of  writing the Lorentz equation
in terms of the kinetic or of the canonical momentum, respectively.
The fact that the field $a^\mu$ turns out to be
 a pure gauge (see eqs.~(\ref{0f}) and (\ref{a2})) results ultimately
from the assumption that
 the trajectories of the plasma particles are not deviated
by the interaction with the gauge fields. In particular,
 $a^\mu$  can be set equal  to zero by an appropriate
choice of the gauge, specifically, by choosing the gauge $v\cdot A=0$
(see eq.~(\ref{a1})). Then,  $W^\mu=-
A^\mu$, as can be seen directly on eq.~(\ref{eqw}).
Note, however, that for practical purposes, such a
 gauge is of no use, since the calculation of the induced current ---
the quantity one is primarily
 interested in --- involves an integration over {\it all}
 directions of ${\bf v}$ (see eq.~(\ref{j1})).

Quite remarkably, the gauge structure that we have
 exhibited completely determines the underlying dynamics.
This explains why most of the dynamical information could have been
 reconstructed from arguments based
on gauge invariance only\cite{TW90}. To see this, we note that
 the dynamical equation (\ref{eqa}) for $a^\mu= A^\mu + W^\mu$
follows  from eq.~(\ref{0f}),  once
 the condition (\ref{trans}) is taken into account.
 Indeed, by multiplying eq.~(\ref{0f}) by $v^\nu$, we obtain
\beq\label{A}
 (v\cdot\del)a_\mu-v^\nu(\del_\mu a_\nu)+ig[v\cdot a, a_\mu] =0.\eeq
Using the fact that $v^\mu$ is a constant vector, together with the
property  $v\cdot a=v\cdot A$ (which follows from
 eqs.~(\ref{trans}) and (\ref{defa})),
we can write this equation in the form (\ref{eqa}).
To the extent that one is interested only in the calculation of the induced
current --- which, according to eq.~(\ref{j1}), depends on $W^0=a^0-A^0$ ---
only the component $\mu =0$ of eq.~(\ref{eqa}) is needed. Explicitly,
\beq\label{e0}
 (v\cdot\del)a_0\,-\,\del_0(v\cdot A)\,+\,ig\,[v\cdot A, a_0] =0.\eeq
This is, within minor notational changes, the equation which has been
 obtained by Taylor and Wong as the condition of
gauge invariance of the effective action \cite{TW90}.
This is  discussed in detail in the next section.

\setcounter{equation}{0}
\section{Relation to previous works}

As we have seen, in the computation of the induced current,
 the important gauge variables are $A^0$ and $v\cdot A =
A^0-{\bf v}\cdot {\bf A}$. (Recall that
$a^0$ is a functional of $v\cdot A$ only, as implied by eq.~(\ref{e0}).)
Alternatively, and
in order to facilitate the comparaison with previous works, it
 is convenient to use the variables
 $A_+\equiv v\cdot A$ and $A_-\equiv v^\prime\cdot A$,
where $v^\mu=(1, {\bf v})$ and $v^{\prime\,\mu}=(1, -{\bf v})$.
We also set $\del_+\equiv v\cdot\del$ and $\del_-\equiv v^\prime\cdot\del$.
In these notations, eq.~(\ref{e0}) reads
\beq\label{eqvl}
\del_+\,a_0\,-\,\del_0\,A_+\,+\,ig\,\bigl[A_+,\,a_0\bigr]\,=\,0.\eeq
We show now that this is precisely the condition for the gauge invariance
of the generating functional of the HTL's, denoted by $S[A_\mu]$.
To do this, let us recall that
the induced current is, formally,
 the first derivative of the effective action $S$,
\beq\label{delS}
\frac{\delta S}{\delta A_\mu^a(x)}=
j^\mu_a(x)\,=\,3\,\omega^2_p\int\frac{d\Omega}{4\pi}
\,v^\mu\,W_a^0(x;v).\eeq Here, ``formally'' means that in writing
 eq.~(\ref{delS}) we are renouncing to the retarded conditions for
$A_\mu$ and $j^\mu$. Indeed, eq.~(\ref{delS}) implies
\beq\frac{\delta j^\mu_a(x)}{\delta A_\nu^b(y)}=
\frac{\delta j^\nu_b(y)}{\delta A_\mu^a(x)},\eeq
which can only hold on the space of fields
in which the operator $v\cdot D$ is invertible\cite{TW90,NPB2}. In
this functional space, which we  denote by  ${\cal R}$,  we can integrate
 (\ref{delS}) to get $S[A]$. Setting $W^0=a^0-A^0$, and separating the
 contributions of $a^0$ and $A^0$, we write eq.~(\ref{delS}) as
\beq\label{delS1}
\frac{\delta S}{\delta A_\mu^a(x)}=
\,3\,\omega^2_p\left\{-g^{\mu 0}A^a_0\,+\,\int\frac{d\Omega}{4\pi}
\,v^\mu\,a_a^0[A_+]\right\},\eeq
where we have indicated explicitly that $a^0$ depends only on
$A_+\equiv v\cdot A$.   The effective action  is therefore of the
form: \beq\label{STW}
S\,=\,3\,\omega^2_p\int d^4x\left\{-\frac{1}{2}\,A_a^0A_a^0\,+\,
\int\frac{d\Omega}{4\pi}\,{\cal W}[A_+]\right\},\eeq
where, by construction, the functional ${\cal W}$ satisfies
\beq\label{delW}
\frac{\delta {\cal W}}{\delta A_+^a(x)}\,=\,a^0_a(x;v).\eeq
Consider now  an infinitesimal gauge transformation
\beq\label{GTR} A_\mu\to A_\mu+\delta A_\mu,\qquad\delta A_\mu=\bigl[ D_\mu,
\theta\,\bigr].\eeq Under such a transformation, $S\to S+\delta S$, with \beq
\label{deltaS} \delta S =3\,\omega^2_p\int d^4x
\int\frac{d\Omega}{4\pi}\,\theta^a\left\{\del_0 A_+^a-\del_+ a_0^a
+gf^{abc}A_+^b a_0^c\right\}.\eeq Since $a^a_0$ satisfies
  eq.~(\ref{eqvl}), it follows that $\delta S =0$.

The form of the effective action given by eq.~(\ref{STW}) was first inferred
by Taylor and Wong from an analysis of the HTL's\cite{TW90}.
 By assuming $S$ to be
gauge invariant, they obtained eq.~(\ref{eqvl}) as an equation for
$\delta W/\delta A_+$. And, in fact, our arguments above have been
intentionally developed in a way which makes the relation
to the analysis of Taylor and Wong  obvious.
But, of course, we could directly infer the gauge invariance of
$S$ from the conservation of
the induced current, through a standard, Noether-type,
argument. That is, by using eqs.~(\ref{delS}) and (\ref{GTR}), we can write
\beq\delta S = \int d^4 x\,j^\mu_a\,\delta  A_\mu
=-\int d^4 x\,\theta^a\,\Bigl[D_\mu,\, j^\mu\Bigr]_a,\eeq
where the second equality follows
after an integration by parts (the surface  term has been assumed
to vanish). Because of the conservation law for $j^\mu$, eq.~(\ref{cons}),
the last integral vanishes, and so does  $\delta S$.
The relation between the condition for the  gauge
invariance of the effective action and the dynamical equation for $j^\mu$
 has also been pointed out in Ref. \cite{Liu93}.

The arguments above suggest a different way to compute the induced current,
which has been, in fact,  followed by Jackiw and Nair in Ref. \cite{JN93}.
Starting  with the effective action $S$ of Taylor and Wong,
 eq.~(\ref{STW}), they have constructed
$j^\mu_a$ as the first derivative of $S$, i.e.
  eq.~(\ref{delS1}), with  $a^0_a[A_+]\equiv \delta W/\delta A_+^a$
determined from  the gauge-invariance condition on $S$, eq.~(\ref{eqvl}),
as in the original work of Taylor and Wong. In order to recognize that the
expression for $j^\mu_a$ obtained  in \cite{JN93} is the same as ours, we
 follow
 Ref. \cite{EN92} and replace eq.~(\ref{eqvl}) with an equivalent equation
for the new function $a_-\equiv 2a_0 -A_+$:
\beq\label{eqam}
\del_+\,a_-\,-\,\del_-\,A_+\,+\,ig\,\bigl[A_+,\,a_-
\bigr]\,=\,0.\eeq
Then, the current (\ref{delS1}) becomes
\beq\label{JN}
j^\mu(x)\,=\,3\,\omega^2_p\int\frac{d\Omega}{4\pi}
\,v^\mu\,\bigl(a_0\,-\,A_0\bigr)
\,=\,\frac{3}{2}\,\omega^2_p\int\frac{d\Omega}{4\pi}
\,v^\mu\,\bigl(a_-\,-\,A_-\bigr).\eeq  We recognize in eqs.~(\ref{eqam})
and (\ref{JN}) the expression of $j^\mu$ proposed in Ref. \cite{JN93}.

On the other hand, it can be easily verified
 using eqs.~(\ref{trans}) and (\ref{defa}) that $a_-$ just defined is also
equal to $ v^\prime\cdot a$. This is not
surprising, since eq.~(\ref{eqam}) also follows from eq.~(\ref{eqa}) for
$a^\mu$  after multiplication with $v^\prime_\mu$. Let us further define
$a_+\equiv v\cdot a$. Because of eqs.~(\ref{trans}) and (\ref{defa}), we  have
$a_+\,=\,A_+$ (i.e.,   $v\cdot a=v\cdot A$, as already noticed after
eq.~(\ref{A})), so that eq.~(\ref{eqam}) may be finally written as
\beq\label{eqmp}
\del_+\,a_-\,-\,\del_-\,a_+\,+\,ig\,\bigl[a_+,\,a_-
\bigr]\,=\,0,\eeq
which coincides with the projection of the zero-curvature condition
(\ref{0f}) onto the hyperplane defined by the 4-vectors $v^\mu$
and $v^{\prime\,\mu}$; that is, eq.~(\ref{eqmp}) is the same as
 $v_\mu\,f^{\mu\nu}v^\prime_\nu =0$. If we recall that $a_+= a_0-{\bf v}\cdot
{\bf a}$ and  $a_-= a_0+{\bf v}\cdot {\bf a}$, we see that the gauge structure
defined by eq.~(\ref{eqmp}) involves the temporal ($a_0$)
and the longitudinal (${\bf v}\cdot{\bf a}$)
components of $a_\mu$, but not its two transverse components.
This equation represents the starting
point of the relation to the Chern-Simons theory explored in Refs. \cite{EN92}.

The main difference between the two derivations of $j_\mu$ outlined
above concerns the method used
for solving eq.~(\ref{eqam}) (or, equivalently, eq.~(\ref{eqw}) for $W^0$).
We get  the  solution directly in Minkovski space, where retarded
conditions are easily implemented (see eq.~(\ref{W}) above).
The authors of Ref. \cite{JN93}  solve eq.~(\ref{eqam})
in Euclidian space-time, using technics borrowed from the study
of the Chern-Simons theory. They obtain thus
 $a_-$ as a power series in $A_+$\cite{EN92}. In order to get the
retarded current,  an analytic continuation has to be
 performed, to ensure the correct $i\epsilon$ prescription\cite{JN93}.

The functions $W^\mu$ have been found  useful in the construction
of the energy-momentum tensor for the gauge fields\cite{Weldon93,EMT}, and
also in the construction of a manifestly gauge
invariant expression for the effective action.
 Such an expression was proposed first by Braaten and Pisarski\cite{PB92}.
Their expression reads (with $(\tilde D F)^a\equiv [D, F]^a$)
\beq\label{Sb}
\qquad S&=&\frac{3}{2}\,\omega^2_p\int \frac{d\Omega}{4\pi}\int d^4 x
\int d^4 y \,{\rm Tr} \left [ F_{\mu\lambda}(x)
\langle x|\frac {v^\mu v_\nu}{(v\cdot \tilde D)^2}|y\rangle
 F^{\nu\lambda}(y)\right ],
\eeq and has been rederived since by Frenkel and Taylor [11] (who inferred $S$
from a study of
 the forward scattering amplitudes for quarks and gluons on ``soft''
gauge fields), and by us (in Ref. \cite{NPB2}, we have shown explicitly
that the functional derivative of  (\ref{Sb}) gives
the correct induced current). The functional (\ref{Sb}) is well
defined only on ${\cal R}$, where it coincides with\cite{Weldon93}
\beq\label{SW}
S&=&-\,\frac{3}{2}\,\omega^2_p\int d^4 x\,\int \frac{d\Omega}{4\pi}\,
{\rm Tr}\, W^\mu(x;v)\,W_\mu(x;v).\eeq
By using eqs.~(\ref{defa}) and (\ref{a2}), we can write
\beq\label{W1}W^\mu(x;v)\,=\,\frac{i}{g}\,\Bigl(D^\mu V\Bigr)\,V^{-1}.\eeq
Thus, an alternative expression for $S$ is
\beq\label{ST}
S&=&-\,\frac{3}{2}\,\frac{\omega^2_p}{g^2}
\int d^4 x\,\int \frac{d\Omega}{4\pi}\,
{\rm Tr}\, \Bigl(D_\mu V\Bigr)\,\Bigl(D^\mu V\Bigr)^\dagger,\eeq
 which also appears in Ref. \cite{BFT93}.

Let us finally consider the energy ${\cal E}$ of a gauge field configuration
 in the plasma. Different approaches has been proposed recently
 to compute this
quantity\cite{Weldon93,BFT93,N93,EMT}, and the various results obtained
are not  related in an obvious way.
 In Ref. \cite{EMT}, we have calculated ${\cal E}$ by integrating
 the field equations of motion (\ref{ava}).
One of our  results may be written in
the form ${\cal E}={\cal E}_{YM}+{\cal E}_{ind}$, where ${\cal E}_{YM}$ is
the standard Yang-Mills contribution,
\beq\label{enloc}
{\cal E}_{YM}\equiv \int d^3 x\,
\frac{1}{2}\Bigl({\bf E}^a(x)\cdot{\bf E}^a(x)\,+\,
{\bf B}^a(x)\cdot{\bf B}^a(x)\Bigr),\eeq
($B^i_a(x)\equiv -(1/2)\epsilon^{ijk}\,F^{jk}_a(x)$), while
\beq\label{ennon}
{\cal E}_{ind}\equiv {3}
\omega^2_p\int d^3 x\int\frac{d\Omega}{4\pi}\,{\rm Tr}\,W^0(x;v)\,W^0(x;v),\eeq
may be interpreted as the {\it polarization energy} of the plasma, that is,
the energy transferred by the gauge fields to the plasma
constituents\cite{EMT}. According to these expressions, the field energy is
 obviously positive. A similar conclusion has been reached by
Nair\cite{N93}, who derived a Hamiltonian for the soft gauge fields
by exploiting the analogy with the Chern-Simons theory. The resulting
energy functional is positive when evaluated for gauge fields which
satisfy Gauss's law. Nair's expression for ${\cal E}$ reads
 ${\cal E}={\cal E}_{YM}+{\cal E}_{N}$, with
\beq\label{enN}
{\cal E}_{N}\equiv \frac{3}{4}\,\frac{\omega^2_p}{g^2}
\int d^3 x\int\frac{d\Omega}{4\pi}\,{\rm Tr}\,\left\{\Bigl[D_0,\,G\Bigr]
\Bigl[D_0,\,G^\dagger\Bigr]\,+\,\Bigl[{\bf v}\cdot {\bf D},\,G\Bigr]
\Bigl[{\bf v}\cdot {\bf D},\,G^\dagger\Bigr]\right\},\eeq
and involves the auxiliary matrix field $G(x,{\bf v})\in SU(N)$,
with the property  $G(x,{\bf v})= G^\dagger(x, -{\bf v})$.

We show now  that, in spite of formal differences,
the energy {\it densities} in eqs.~(\ref{ennon}) and (\ref{enN}) are, in fact,
the same. To do that, we first
 replace $D_0=(D_-+D_+)/2$ and ${\bf v}\cdot {\bf D}=(D_--D_+)/2$
(where $D_+\equiv v\cdot D$ and $D_-\equiv v^\prime\cdot D$). We thus get,
instead of (\ref{enN}),
\beq\label{enN1}
{\cal E}_{N}\equiv \frac{3}{8}\,\frac{\omega^2_p}{g^2}
\int d^3 x\int\frac{d\Omega}{4\pi}\,{\rm Tr}\,\left\{\Bigl[D_-,\,G\Bigr]
\Bigl[D_-,\,G^\dagger\Bigr]\,+\,\Bigl[D_+,\,G\Bigr]
\Bigl[ D_+,\,G^\dagger\Bigr]\right\}.\eeq
By changing ${\bf v}\to -{\bf v}$ in the second term inside the curly braces,
we see that it gives the same contribution as the first one after  angular
integration. Furthermore, we note that the equations
of motion for  $G(x,{\bf v})$, as written down in Ref. \cite{N93},
are implicitly solved by
\beq\label{aG}
a_-=G^{-1}\,A_-\,G\,-\,\frac{i}{g}\,G^{-1}\,\del_-\,G,\eeq
which relates $G$ to our function $a_-\equiv v^\prime\cdot a$.
Writing this equation in the form
\beq a_--A_-=-\frac{i}{g}\,G^{-1}\,\Bigl[D_-,\,G\Bigr]=
\frac{i}{g}\,\Bigl[D_-,\,G^\dagger\Bigr]\,G,\eeq
we conclude that eq.~(\ref{enN1}) is equivalent to
\beq\label{enN2}
{\cal E}_{N}\equiv \frac{3}{4}\,{\omega^2_p}
\int d^3 x\int\frac{d\Omega}{4\pi}\,{\rm Tr}\,\Bigl\{
\bigl(a_--A_-\bigr)\,\bigl(a_-- A_-\bigr)\Bigr\},\eeq
where any direct reference to the field $G$ has disappeared. It is then
sufficient to recall that $a_--A_-=2W^0$ to see that ${\cal E}_N$, as given
by eq.~(\ref{enN2}), is identical to our expression (\ref{ennon}) for
${\cal E}_{ind}$.
Remark that in the arguments above we did not need the explicit expression
of $G$. However, if one recalls that $a_-=(i/g)(\del_-V)V^{-1}$ (see
eq.~(\ref{a2})), with $V(x;v)$ defined by (\ref{defV}), one can easily
verify that  eq.~(\ref{aG}) is solved by $G(x,{\bf v})\,=\,V(x;v^\prime)\,
V^\dagger(x; v)$.

\section{Conclusions}

The classical features of the HTL's reflect the long-wavelength, collective
behaviour of the plasma particles, namely, their average motion over
distances which are large compared to their mean separation. The coherent
motion of the particles is described, to leading order in $g$, by density
matrices obeying simple  kinetic equations. In the abelian case, the
kinetic equation is the linearized Vlasov equation and the elementary
dynamics is essentially that of the Lorentz equation which governs
 the motion of a classical charged
particle in an electromagnetic field. In the non-abelian case, the
distribution functions are matrices in color space and their time dependence
involves in addition a ``precession'' in color space.
 We have shown that a remarkably simple gauge
structure emerges from our kinetic equations once
one assumes that the plasma particles are not
significantly deviated by the background field. Remarkably, this gauge
structure determines completely
the collective dynamics of the particles at the order of the
 HTL's approximation. This explains why the requirement of gauge symmetry
was, in fact, a sufficient condition to obtain the effective action generating
the HTL's\cite{TW90}, as well as the related field equations\cite{EN92,JN93}.

\vspace{1cm}
\noindent
{\bf Acknowledgements}\\
\noindent We would like to thank V.P. Nair for useful correspondance
concerning his work on the energy of gauge field configurations\cite{N93}.

\end{document}